\documentclass[a4paper]{article}

\usepackage{amssymb}
\usepackage{amsfonts}
\usepackage{amsmath}
\usepackage{graphicx}
\usepackage{color}
\usepackage{comment}
\usepackage{hyperref}
\hypersetup{backref,
            pagebackref=true,
            ps2pdf,
            bookmarks=true,
            bookmarksnumbered=true,
            pdfauthor=McCabe Wydro,
            colorlinks=true,
            linkcolor=blue}

\title{Critical Excitation Spectrum of Quantum Chain\\
With A Local 3-Spin Coupling}

\author{
\textrm{JOHN F. McCABE}\\%
 \textit{ 2331 Gales Court, Scotch Plains, NJ 07076, USA}\\
jfmccabe@lycos.com
\\  \\
\textrm{TOMASZ WYDRO}\\%
\textit{ Statistical Physics Group, P2M Dpt, Institut Jean Lamour}\\
  \textit{Nancy Universit\'e, Universit\'{e} Paul Verlaine - Metz}\\
  \textit{BP 70239, 54506 Vandoeuvre-les-Nancy Cedex, France}\\
wydro@lpm.u-nancy.fr
 }

\begin{document}

\maketitle

\begin{abstract}
This article reports a measurement of the low-energy excitation spectrum along the critical
line for a quantum spin chain having a local interaction between three Ising spins and
longitudinal and transverse magnetic fields.  The measured excitation spectrum agrees with
that predicted by the (D$_4$, A$_4$) conformal minimal model under a nontrivial
correspondence between translations at the critical line and discrete lattice translations.
Under this correspondence, the measurements confirm a prediction that the critical line of
this quantum spin chain and the critical point of the 2D 3-state Potts model are in the
same universality class.
\end{abstract}

\begin{center}
\bigskip
\end{center}

The solution of the two-dimensional (2D) 8-vertex model was a breakthrough in the study of
lattice models in statistical mechanics \cite{Baxter}.  Soon after the publication of this
solution, the 2D 8-vertex model was shown to be equivalent to a model with local
interactions between four Ising spins \cite{Wu}. This equivalence stimulated studies of
other models with local interactions between three and four Ising spins.  One such 2D model
is described by the Hamiltonian, $H$:
\begin{equation}
H=-\sum_{i,j}(J_1 \cdot S_{i,j}S_{i+1,j}S_{i+2,j}+ J_2\cdot S_{i,j}S_{i,j+1}+h S_{i,j})
\textrm{ .} \label{H3spin}
\end{equation}
In Eq. (\ref{H3spin}), $J_1$ and $J_2$ are local 3-spin couplings in the respective "x" and
"y" directions, and $h$ is a magnetic field.  For $J_1 \neq J_2$, the couplings between the
Ising spins are anisotropic.  In the extreme anisotropic limit, a 2D model's critical
behavior is also described by a quantum spin chain \cite{Fradk78}. The quantum spin chain
for the 2D model of Eq. (\ref{H3spin}) has a local interaction between three Ising spins
and interactions with a magnetic field \cite{Penson}. This quantum spin chain is described
by a Hamiltonian, $\widehat{H}$, given by:
\begin{equation}
\widehat{H}=-\sum_{i}\sigma^z_{i}\sigma^z_{i+1}\sigma^z_{i+2} - h_l\sum_{i}\sigma^z_{i} -
h_t\sum_{i}\sigma^x_{i} \textrm{ .} \label{H3chain}
\end{equation}
In Eq. (\ref{H3chain}), $\sigma^x_{i}$, $\sigma^y_{i}$, and $\sigma^z_{i}$ are the $x$,
$y$, and $z$ Pauli matrices at site $i$, and $h_l$ and $h_t$ are the respective
longitudinal and transverse components of the magnetic field.

In the $h_l$ and $h_t$ parameter plane, the quantum spin chain of Eq. (\ref{H3chain}) has a
rich phase structure for special values of the length, L, of the chain. For the special
values, L = 3N with N being any positive integer \cite{Penson}.  For these lengths, L, the
quantum spin chain has a 3-fold degenerate ground state in a region of the $h_l$ and $h_t$
plane.  The region is bounded by a segment of the $h_l$-axis for which $-3 \leq h_l \leq
0$, a segment of the $h_t$-axis for which $0 \leq h_t \leq 1$, and a curved line segment
connecting the $(h_l, h_t)$ points (-3, 0) and (0, 1).

Along the curved line segment, there is evidence that the quantum spin chain of
Eq.(\ref{H3chain}) is critical \cite{Penson}. Indeed, Penson et al provided evidence
identifying this critical line corresponds to a 2D conformal field theory (CFT) \cite{BPZ}.
The CFT has a central charge, $c$, of 0.8, a magnetic exponent $y_h$ of about 1.874, and a
$\nu$ exponent of about 0.839.  These measured values are close to the values $c=4/5$, $y_h
= 28/15$, and $\nu = 5/6$ at the critical point of the 2D 3-state Potts model. For that
reason, Penson et al concluded that the critical line of the quantum spin chain of Eq.
(\ref{H3chain}) is in the universality class of the critical point of the 2D 3-state Potts
model.

The critical point of the 2D 3-state Potts model is described by the (D$_4$, A$_4$)
conformal minimal model of the ADE classification \cite{ADE1}. The identification of a
critical line of the quantum spin chain of Eq. (\ref{H3chain}) with the (D$_4$, A$_4$) CFT
is surprising, because the (D$_4$, A$_4$) CFT has the $Z_3$ symmetry of the 3-state Potts
model  \cite{MCCABE}.  While the 3-state Potts model has a 3-state spin that causes the
$Z_3$ symmetry, there is only a 2-state Ising spin in quantum spin chain of Eq.
(\ref{H3chain}).  For this reason, the quantum spin chain of Eq. (\ref{H3chain}) cannot
have a $Z_3$ internal symmetry.  Due to the absence of such an internal symmetry, the
identification of the curved critical line of the quantum spin chain of Eq. (\ref{H3chain})
with the (D$_4$, A$_4$) CFT is worthy of further study.

This article evaluates the spectrum of low-lying excitations at the curved critical line of
the quantum spin chain of Eq. (\ref{H3chain}) when periodic boundary conditions are
imposed.  In particular, both energies and momenta of excitations are determined for chains
of triple integer lengths, i.e., L = 3N with N being an integer.

In this article, values of $h_l(L)$ and $h_t(L)$ are defined through the phenomenological
renormalization group (PRG) \cite{Derrida}.  When evaluated at PRG values of $h_l(L)$ and
$h_t(L)$, physical quantities scale with the size of the system.  The scaling behavior
enables the extraction of the values of such physical properties at critical points and
critical lines. For the quantum spin chain of Eq. (\ref{H3chain}), the equation for the PRG
was used in the form:
\begin{equation}
(L-3)\cdot m(h_l(L), h_t(L), L-3)= L\cdot m(h_l(L), h_t(L), L)\textrm{ .} \label{PRG}
\end{equation}
In Eq. (\ref{PRG}), m($h_l$, $h_t$, L) is the energy gap, which is defined by $m(h_l,
h_t,L) = \left[E_{1}(h_l, h_t, L) - E_{0}(h_l, h_t, L)\right]$.  Here, $E_0(h_l, h_t, L)$
and $E_1(h_l, h_t, L)$ are the energies of the ground state "0" and the first excited state
"1" of the quantum spin chain of length $L$.  For our measurements, $h_l(L)$ was
arbitrarily set to -1, and $h_t(L)$ was determined by solving Eq. (\ref{PRG}).  For this
selection of $h_l(L)$, the PRG values of $h_t(L)$ were found to be 1.1055, 1.05425,
1.04584, and 1.04277 for L equal to 6, 9, 12, and 15, respectively.  The corresponding
sequence of ($h_l(L)$, $h_t(L)$) points seems to converge to a point on the curved critical
line of reference \cite{Penson} thereby showing that our methods reproduce properties at
the curved critical line of Penson et al.

Since the Hamiltonian, $\widehat{H}$, of Eq. (\ref{H3chain}) is invariant under
translations by a single lattice site, the Hamiltonian, $\widehat{H}$, was separately
diagonalized over eigenspaces of the operator $T_{1ls}$, which generates such translations.
Each eigenspace of $T_{1ls}$, corresponding to a momentum eigenvalue p, includes a set of
one or more one-dimensional (1D) Bloch states, i.e., $\Phi_{p,i}(x)$'s.  Under a
translation by a single-lattice site, such a Block state satisfies: $\Phi_{p, i}(x + 1) =
exp[(2\pi ip/L)]\Phi_{p,i}(x)$ where $exp[(2\pi ip/L)$ is the eigenvalue of one
lattice-site translation operator $T_{1ls}$. Here, "$p$" is a modular integer of (-L/2,
+L/2), which defines the discrete lattice momentum, i.e., an eigenvalue, $P_{1ls}$,
associated with $T_{1ls}$, and $i$ is an integer indexing the one or more state with the
same eigenvalue $P_{1ls}$.  By using 1D Block states, both energies and the discrete
lattice momenta may be measured and then, compared with the predictions of the (D$_4$,
A$_4$) CFT.

At criticality, CFT predicts the scaling of excitation energies and momenta with the chain
length, L.  For an excitation state "i", the normalized excitation energy, $E_i$, will be
of the form \cite{Cardy1-84}:
\begin{equation}
E_i = [E_{i}(L)-E_{0}(L)]/E_0(L) =\frac{(\Delta_i + {\bar\Delta_i})- (\Delta_0
+{\bar\Delta_0}) }{(\Delta_0 + {\bar\Delta_0})}\textrm{ ,} \label{gaps scaling}
\end{equation}
and the normalized excitation momentum, $P_i$, will be of the form:
\begin{equation}
p_{i}(L)= (\Delta_i - {\bar\Delta_i})\textrm{ .} \label{momenta scaling}
\end{equation}
Here, $\Delta_{i}$ and ${\bar\Delta_{i}}$ are the respective left and right conformal
dimensions of the excitation "i".  In Eq. (\ref{gaps scaling}), the excitation energy,
$E_i$, has been normalized with respect to ground state energy to remove the leading
dependence on the chain length, L.  In Eq. (\ref{momenta scaling}), the excitation
momentum, $P_i$, has been normalized to remove the dependency on the chain length, L. These
normalizations simplify comparisons between quantum spin chains of different length, L.

For the conformal minimal models, the ADE classification provides the excitation spectrum
\cite{ADE1,Rocha-Caridi}. For the $(D_{4}, A_{4})$ CFT, the ADE classification predicts
that a state "i" will have a left conformal dimension $\Delta_{i}$, a right conformal
dimension ${\bar\Delta_{i}}$, a normalized excitation energy $E_i$, a normalized excitation
momentum $P_i$, and a degeneracy g(i) as shown in Table 1.
\bigskip
\begin{center}
\begin{tabular}{|c||c|c|c|c|c|c|}
\multicolumn{7}{c}{Table 1: Low-lying States of the $(D_{4}, A_{4})$ CFT}\\
\hline
  &  &  &  &  &  & \\
State i & $\Delta_i$ & ${\bar\Delta_i}$ & $E_{i}$ & $P_i$ & g(i) & $Z_3$ representation\\
\hline
A &    0  &      0 & 0    &  0  & 1 & 1 singlet\\
\hline
B & 1/15  &   1/15 & 1    &  0  & 2 & 1 doublet\\
\hline
C &  2/5  &    2/5 & 6    &  0  & 1 & 1 singlet\\
\hline
D & 16/15 &  1/15  & 8.5  &  1  & 2 & 1 doublet\\
\hline
E &  1/15 & 16/15  & 8.5  & -1  & 2 & 1 doublet\\
\hline
F &   2/3 &  2/3   & 10   &  0  & 2 & 1 doublet\\
\hline
G &   7/5 &  2/5   & 13.5 &  1  & 2 & 2 singlets\\
\hline
H &   2/5 &  7/5   & 13.5 & -1  & 2 & 2 singlets\\
\hline
\end{tabular}
\end{center}
\bigskip

By solving PRG Eq. (\ref{PRG}), the low-lying excitation spectrum of the quantum spin chain
of Eq. (\ref{H3chain}) was measured for chains having lengths, L, of 9, 12, and 15.  For
each value of the lattice momentum, $P_{1ls}$, the excitation energies were determined with
a commercial linear algebra package of Maple$_{TM}$.  For the PRG solution with L = 15,
Figure (1) plots the measured excitation energies as a function of lattice momenta, i.e.
$P_{1ls}$'s.

\begin{figure} [h t p]
\centering
    \includegraphics[width=12cm]{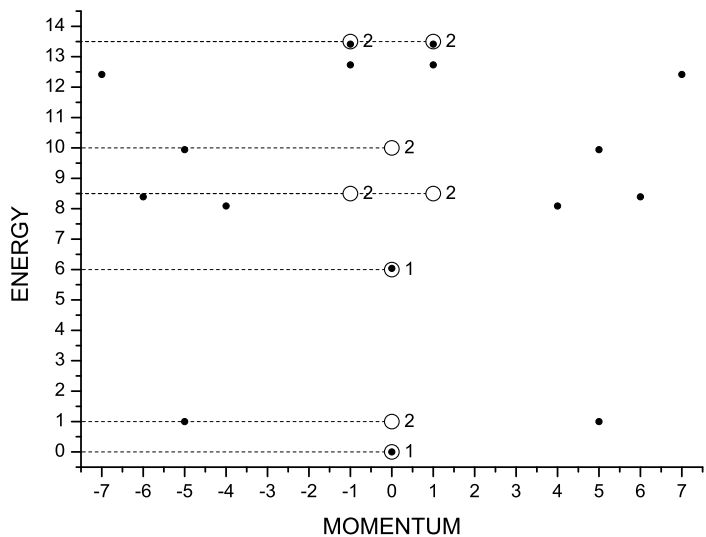}
\caption{Measured energies and momenta, i.e. $P_{1ls}$'s, of low lying states of the 3-spin
quantum chain of length 15 (black dots) and predicted energies and momenta of the  $(D_{4},
A_{4})$ CFT (empty circles).} \label{fig: T1}
 \end{figure}

An inspection of Figure (1) shows that measured excitation energies and degeneracies agree
reasonably well with predictions of the $(D_{4}, A_{4})$ CFT for the states A - H of Table
1. But, such an inspection also shows clear differences between the observed lattice
momenta, i.e., the $P_{1ls}$'s, and the predicted normalized momenta as shown in Table 1.
Since the predicted momenta do not correspond to the $P_{1ls}$'s associated with the single
lattice-site translation operator $T_{1ls}$, a single lattice-site translation cannot be
the discrete version of a translation at the curved critical line of the quantum spin chain
of Eq. (\ref{H3chain}).  Based on the differences in the momentum spectra, another discrete
symmetry of the quantum spin chain must provide the discrete version of a translation at
the curved critical line.

The form of such translations is elucidated by Figure (2), which provides a second plot of
the spectra for the PRG solution of Eq. (\ref{PRG}) when L = 15.  The second plot shows
measured excitation energies as a function of 1/3 of the momentum, $P_{3ls}$, associated
with a translation by three lattice-sites.  Here, the measured momenta are $P_{3ls}$/3,
which includes a division by three to produce integers as in Table 1.

\begin{figure} [h t p]
\centering
    \includegraphics[width=12cm]{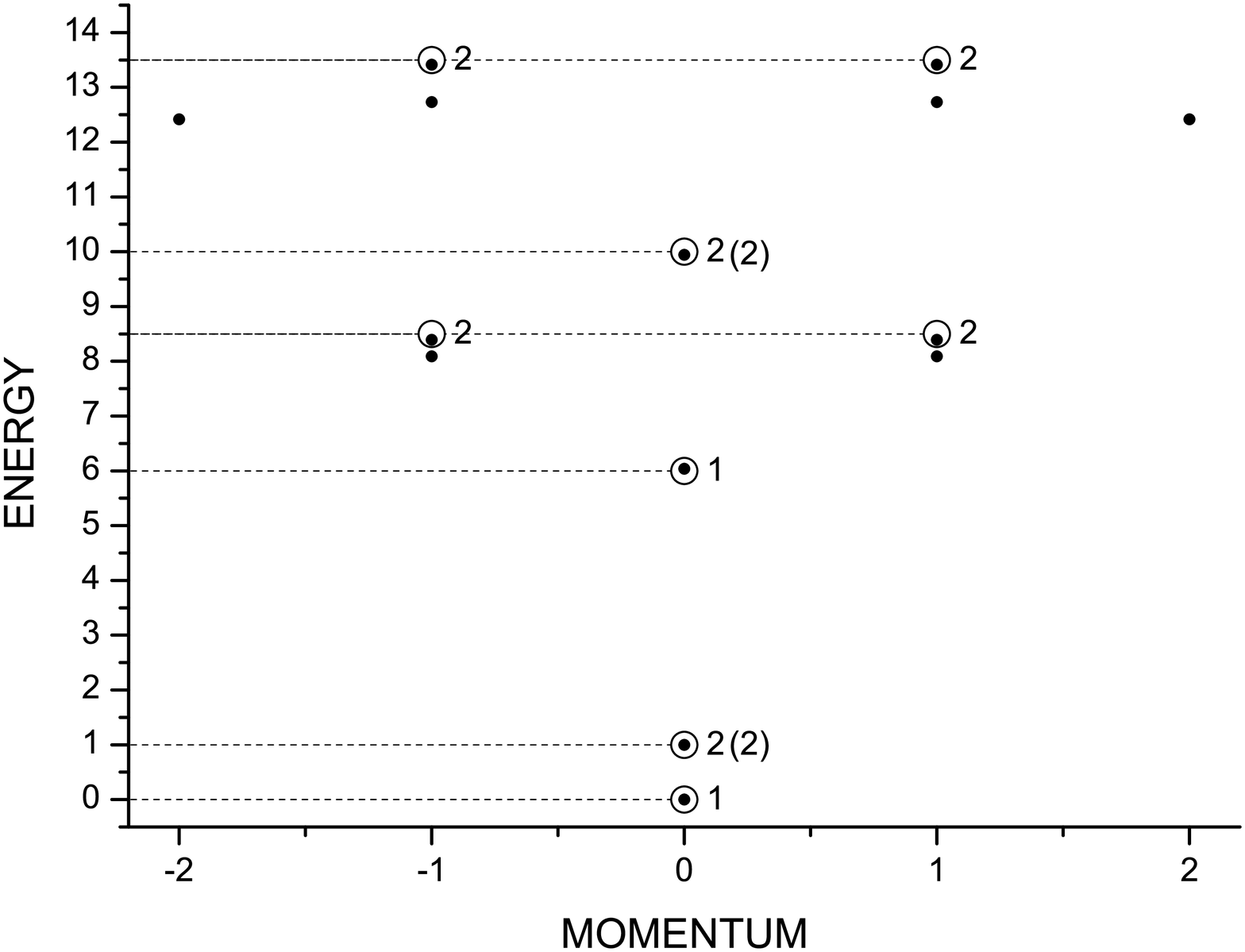}
\caption{Measured energies and folded momenta, i.e., $P_{3ls}$'s, of low-lying states of
the 3-spin quantum chain of length 15 (black dots) and predicted energies and momenta of
the $(D_{4}, A_{4})$ CFT (empty circles). } \label{fig: T3}
 \end{figure}

For a state "i", the lattice momentum $P_{3ls}(i)$ solves:
\begin{equation}
exp^{(2\pi iP_{3ls}(i)/L)} = exp^{(6\pi i P_{11s}(i)/L)} \label{LP}
\end{equation}
where $P_{11s}(i)$ is the momentum associated with a translation by a single lattice-site.
Since values of discrete momenta are required to be in (-L/2, +L/2) for a chain of length
L, Eq. (\ref{LP}) implies that $P_{3ls}(i) = (P_{11s}(i) + N(i))/3$ where N(i) is an
integer that is fixed by requiring that $P_{3ls}$(i) belong to (-L/2, +L/2).  The last
equation produces the $P_{3ls}$(i) values by performing a folding operation on $P_{11s}$(i)
values.  The folding operation qualitatively changes the excitation spectrum.  After the
folding operation, the measured energies and momenta agree well with the predicted values
of the $(D_{4}, A_{4})$ CFT.  Thus, for the quantum spin chain of Eq. (\ref{H3chain}), a
translation by three lattice-sites is the discrete version of a translation at the curved
critical line.

Figure (2) also shows that the D and E states are not exact energy doublet states. In
contrast, the $(D_{4}, A_{4})$ CFT predicts that these states are doublets under a discrete
$Z_2$-symmetry as shown in Table 1.  Thus, the discrete ${Z_3}$-symmetry of the $(D_{4},
A_{4})$ CFT is broken at the PRG solutions.

For L = 9, 12, and 15, the measured excitation energies of the quantum spin chain Eq.
(\ref{H3chain}) were also fitted to scaling forms to evaluate scaling behaviors as $L
\to\infty$. For each state "i", the measured excitation energy, $E_i(L)$, was fit to a
scaling equation $ E_i(L)= E_i(\infty) + x_i/L^y$ where the last term is the leading
finite-length correction.  The parameters $x_i$ and y vary with the state "i", because
scaling corrections depend on the states' conformal dimensions.  For the state "i",
$E_i(\infty)$ is the limit of the normalized excitation energy when $L \to\infty$.  The
fitted values of the  $E_i(\infty)$'s are shown in Table 2.
\begin{center}
\begin{tabular}{|c||c|c|c|}
\multicolumn{4}{c}{Table 2: Fitted Excitation Energies for $L\to\infty$}
\\
\hline  &  &  &  \\
State & g(i) & $P_{3ls}(i)$ & Fitted $E_i$ \\
\hline
 C  & 2  &  0  &  1 \\
\hline
 C  & 1  &  0  &  6.1 \\
\hline
 D & 1  &  1   &  8.5 \\
\hline
 D & 1  &  1   &  9.0 \\
\hline
 E & 1  & -1   &  8.5 \\
\hline
 E & 1  & -1   &  9.0 \\
\hline
 F  & 2  &  0  & 10.4 \\
\hline
 G & 1  &  1   & 13.5 \\
\hline
 G & 1  &  1   & 14.5 \\
\hline
 H & 2  & -1   & 13.5 \\
\hline
 H & 2  & -1   & 14.5 \\
\hline
\end{tabular}\label{tab:Long-Spectra}
\end{center}
The fitted excitation energies of Table 2 and the predicted excitation energies of Table 1
agree well for normalized energies of about 12 or less, e.g., errors are less than about 6
percent.  For higher energies, measurements are needed on quantum spin chains with longer
lengths, L.
For example, in Figure 2, the states with energies of about 12.3 and momenta of $\pm 2$
were determined to converge to states with energies higher than 15 as $L \to\infty$ by
comparing solutions of the PRG for L = 9, 12, and 15.

In conclusion, our measurements of the excitation energies and lattice momenta of low-lying
excitations indicate that the curved critical line of the 3-spin quantum chain is in the
same universality class as the critical point of the 2D 3-state Potts model, i.e., both
being described by the (D$_4$, A$_4$) CFT.  But, the relationship between discrete lattice
translations and translations of the critical theory is nontrivial.  Also, the discrete
$Z_3$-symmetry of the conformal field theory is only recovered as the chain length becomes
infinite.

\end{document}